\documentclass[aps,prl,amsmath,floats,floatfix, twocolumn, article, superscriptaddress,nofootinbib,showpacs]{revtex4-1}

\usepackage[colorlinks, pdfborder={0 0 0}, plainpages=false]{hyperref}
\usepackage{graphicx}
\usepackage{grffile}
\usepackage{subfigure}
\usepackage{xspace}
\usepackage[usenames,dvipsnames]{color}
\usepackage{amssymb}
\usepackage[normalem]{ulem} 
\usepackage{letltxmacro}
\usepackage{amsfonts}
\usepackage{amsmath}
\usepackage{amssymb}
\usepackage{bm}
\usepackage{dcolumn}
\usepackage{epsfig}
\usepackage{graphicx}
\usepackage{graphics}
\usepackage[latin1]{inputenc}
\usepackage{latexsym}
\usepackage{rotating}
\usepackage{hyperref}
\usepackage[usenames]{color}
\usepackage{float}
\usepackage{xspace} 
\usepackage{mathrsfs}
\usepackage{subfigure}
\usepackage{multirow}

\def\prd{{\it Phys. Rev.} D~}

\def\lesssim{\mathrel{\hbox{\rlap{\hbox{\lower4pt\hbox{$\sim$}}}\hbox{$<$}}}}
\def\gtrsim{\mathrel{\hbox{\rlap{\hbox{\lower4pt\hbox{$\sim$}}}\hbox{$>$}}}}
\def\alt{\mathrel{\hbox{\rlap{\hbox{\lower4pt\hbox{$\sim$}}}\hbox{$<$}}}}
\def\agt{\mathrel{\hbox{\rlap{\hbox{\lower4pt\hbox{$\sim$}}}\hbox{$>$}}}}

\def\gta{\ifmmode {\mathbin{\lower 3pt\hbox   
			{$\,\rlap{\raise 5pt\hbox{$\char'076$}}\mathchar"7218\,$}}}
	\else {${\mathbin{\lower 3pt\hbox
				{$\rlap{\raise 5pt\hbox{$\char'076$}}\mathchar"7218\,$}}}
		$}\fi}
\def\lta{\ifmmode {\,\mathbin{\lower 3pt\hbox   
			{$\,\rlap{\raise 5pt\hbox{$\char'074$}}\mathchar"7218\,$}}}
	\else {${\mathbin{\lower 3pt\hbox
				{$\rlap{\raise 5pt\hbox{$\char'074$}}\mathchar"7218\,$}}}
		$}\fi}

\newcommand{\msun}{{\rm M}_{\odot}}
\newcommand{\beq}{\begin{equation}}
\newcommand{\eeq}{\end{equation}}
\newcommand{\bea}{\begin{eqnarray}}
\newcommand{\eea}{\end{eqnarray}}

\newcommand{\NCSA}{\affiliation{NCSA, University of Illinois at Urbana-Champaign, Urbana, IL, 61801, USA}}
\newcommand{\Stat}{\affiliation{Department of Statistics, University of Illinois at Urbana-Champaign, Urbana, IL, 61801, USA}}
\newcommand{\ANCSA}{\affiliation{Department of Astronomy, University of Illinois at Urbana-Champaign, Urbana, IL, 61801, USA}}

\newcommand{\ECE}{\affiliation{Department of Electrical and Computer Engineering, University of Illinois at Urbana-Champaign, Urbana, IL, 61801, USA}}

\begin{document}

\title{Denoising Gravitational Waves using Deep Learning \\with Recurrent Denoising Autoencoders}

\author{Hongyu Shen}\NCSA\Stat 
\author{Daniel George}\NCSA\ANCSA
\author{E. A. Huerta}\NCSA
\author{Zhizhen Zhao}\NCSA\ECE

	\begin{abstract}
		Gravitational wave astronomy is a rapidly growing field of modern astrophysics, with observations being made frequently by the LIGO detectors. Gravitational wave signals are often extremely weak and the data from the detectors, such as LIGO, is contaminated with non-Gaussian and non-stationary noise, often containing transient disturbances which can obscure real signals. Traditional denoising methods, such as principal component analysis and dictionary learning, are not optimal for dealing with this non-Gaussian noise, especially for low signal-to-noise ratio gravitational wave signals. Furthermore, these methods are computationally expensive on large datasets. To overcome these issues, we apply state-of-the-art signal processing techniques, based on recent groundbreaking advancements in deep learning, to denoise gravitational wave signals embedded either in Gaussian noise or in real LIGO noise. We introduce SMTDAE, a Staired Multi-Timestep Denoising Autoencoder, based on sequence-to-sequence bi-directional Long-Short-Term-Memory recurrent neural networks. We demonstrate the advantages of using our unsupervised deep learning approach and show that, after training only using simulated Gaussian noise, SMTDAE achieves superior recovery performance for gravitational wave signals embedded in real non-Gaussian LIGO noise.

	\end{abstract}
\maketitle	
	\section{Introduction}
	
	The application of machine learning and deep learning techniques have recently driven disruptive advances across many domains in engineering, science, and technology~\cite{DLNature}. The use of these novel methodologies is gaining interest in the gravitational wave (GW) community. Convolutional neural networks were recently applied for the detection and characterization of GW signals in real-time~\cite{2017arXiv170100008G,DNN2}. The use of machine learning algorithms have also been explored to address long-term challenges in GW data analysis for classification of the imprints of instrumental and environmental noise from GW signals~\cite{jade:2015CQGra,jade1:2016,GravitySpy,DeepTransferLearning,DeepTransferNIPS}, and also for waveform modeling~\cite{ENIGMA}. \cite{torres2015split,1409.7888,1602:} introduced a variety of methods to recover GW signals embedded in additive Gaussian noise.
	
	PCA is widely used for dimension reduction and denoising of large datasets~\cite{jolliffe2002principal,anderson1958introduction}. This technique was originally designed for Gaussian data and its extension to non-Gaussian noise is a topic of ongoing research~\cite{jolliffe2002principal}. Dictionary learning~\cite{Mairal:2009:ODL:1553374.1553463,Hawe:2013:SDL:2514950.2515881,NIPS2008_3448} is an unsupervised technique to learn an overcomplete dictionary that contains single-atoms from the data, such that the signals can be described by sparse linear combinations of these atoms~\cite{aharon2006k,sparseRep1}. Exploiting the sparsity is useful for denoising, as discussed in~\cite{sparseRep1,sparseRep2, Mairal:2009:ODL:1553374.1553463,Hawe:2013:SDL:2514950.2515881,NIPS2008_3448}. Given the dictionary atoms, the coefficients are estimated by minimizing an error term and a sparsity term, using a fast iterative shrinkage-thresholding algorithm~\cite{beck2009fast}. 

	Dictionary learning was recently applied to denoise GW signals embedded in Gaussian noise whose peak signal-to-noise ratio (SNR)\footnote{Peak SNR is defined as the peak amplitude of the GW signal divided by the standard deviation of the noise after whitening. We have also reported the optimal matched-filtering SNR (MF SNR)~\cite{saton} alongside the peak SNR in this paper.}\(\sim 1\)~\cite{PhysRevD.94.124040}. This involves learning a group of dictionary atoms from true GW signals, and then reconstructing signals in a similar fashion to PCA, i.e., by combining different atoms with their corresponding weights. However, the drawback is that coefficients are not simply retrieved from projections but learned using $L_1$ minimization. Therefore, denoising a single signal requires running $L_1$ minimization repeatedly, which is a bottleneck that inevitably leads to delays in the analysis. Furthermore, it is still challenging to estimate both the dictionary and the sparse coefficients of the underlying clean signal when the data is contaminated with non-Gaussian noise~\cite{chainais2012towards,giryes2014sparsity}. 

\begin{figure*}
		\begin{center}
			\label{structure}
		\hspace{-.2in}	\subfigure[MTDAE]{\label{MTDAE}\includegraphics[height=0.25\textwidth]{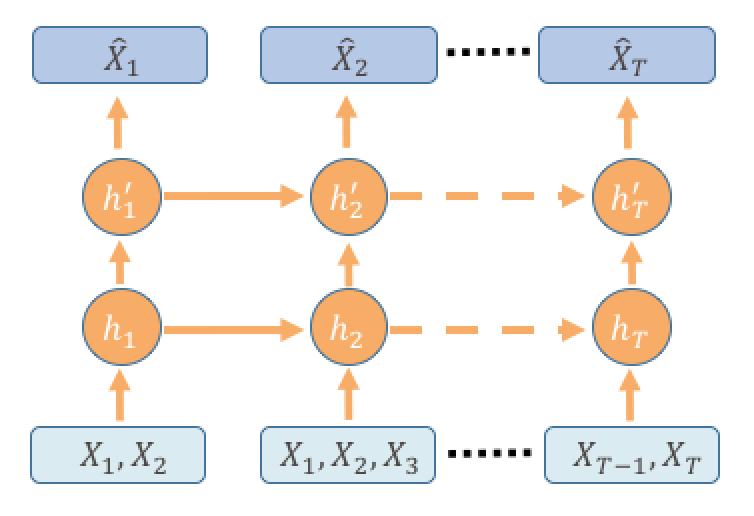}} \enskip
			\subfigure[SMTDAE]{\label{SMTDAE}\includegraphics[height=0.34\textwidth]{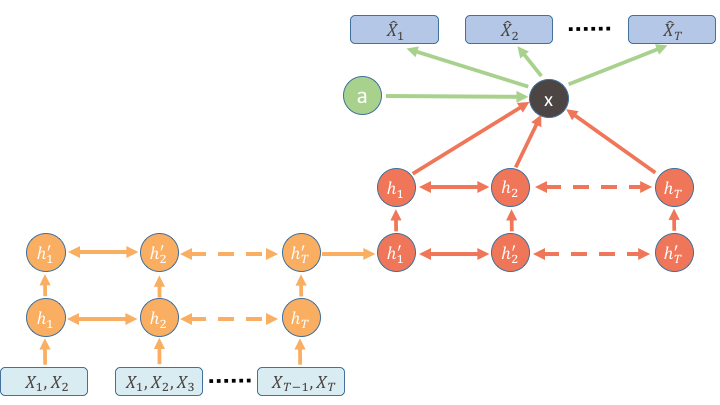}}
			\subfigure[Reference]{\label{REF}\includegraphics[height=0.27\textwidth]{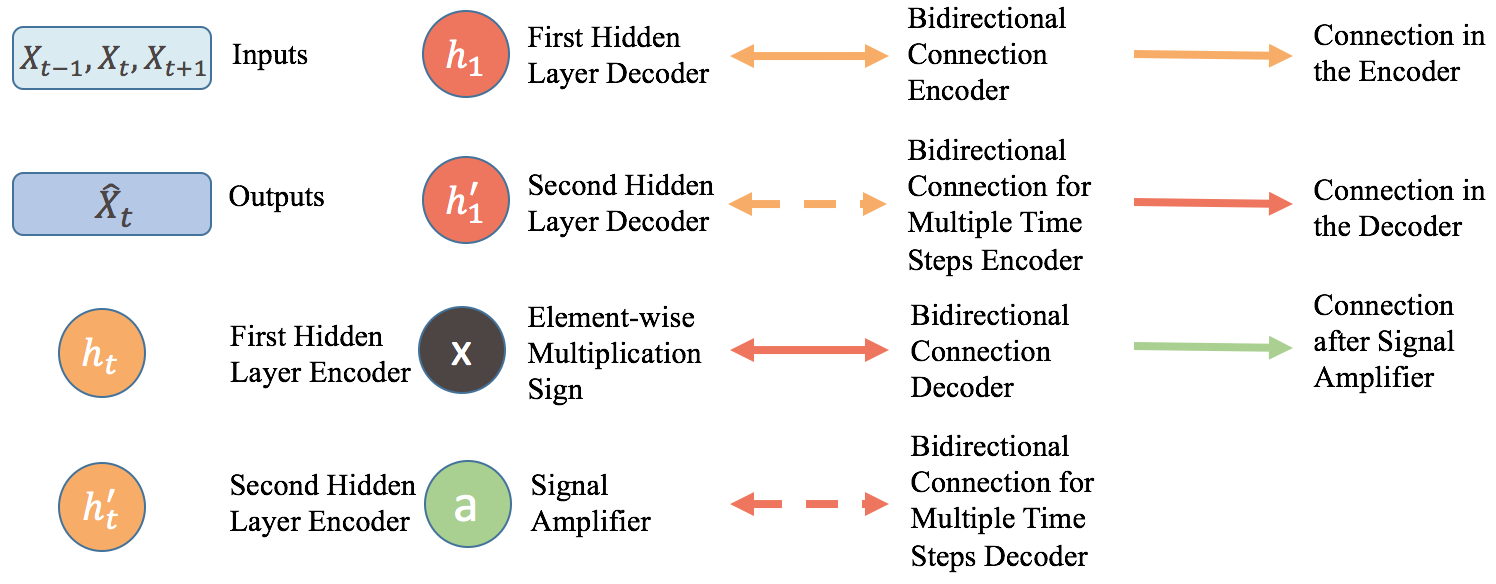}}	
			\caption{\ref{MTDAE} shows the structure of MTDAE. This model passes multiple inputs of a noisy waveform into hidden layers constructed with LSTM cells and outputs a clean version. The timestep in the output is the middle timestep in each of the multiple inputs. \ref{SMTDAE} indicates our proposed SMTDAE structure that uses a sequence-to-sequence model. It differs from MTDAE in that the final state of encoder is passed to the beginning state of decoder in hidden layers. We also include a Signal Amplifier before the output layer in the network to enhance signal reconstruction. The nomenclature is described in~\ref{REF}.}
			\vspace{-5mm}
		\end{center}
	\end{figure*}

	To address the aforementioned challenges, we introduce an unsupervised learning technique using a new model which we call Staired Multi-Timestep Denoising Autoencoder (SMTDAE), that is inspired by the recurrent neural networks (RNNs) used for noise reduction introduced in~\cite{DBLP:conf}. The structure of the SMTDAE model is shown in FIG~\ref{SMTDAE}. RNNs are the state-of-the-art generic models for continuous time-correlated machine learning problems, such as speech recognition/generation~\cite{1303.5778, 1702.07825,1705.10874}, natural language processing/translation~\cite{sutskever2014sequence}, handwriting recognition~\cite{graves2009offline}, etc. A Denoising Autoencoder (DAE) is an unsupervised learning model that takes noisy signals and return the clean signals~\cite{1305.6663,Vincent:2008:ECR:1390156.1390294,Vincent:2010:SDA:1756006.1953039,DBLP:conf}. By combining the advantages of the two models, we demonstrate excellent recovery of weak GW signals injected into real LIGO noise based on the two measurements, Mean Square Error (MSE) and Overlap~\footnote{Overlap is calculated via matched-filtering using the PyCBC library~\cite{Usman:2015kfa} between a denoised waveform and a reference waveform.}~\cite{Canton:2014ena,Usman:2015kfa}. Our results show that SMTDAE outperforms denoising methods based on PCA and dictionary learning using both metrics.	
	
	\section{Methods}
	\label{method}
	
	The noise present in GW detectors is highly non-Gaussian, with a time-varying (non-stationary) power spectral density. Our goal is to extract clean GW signals from the noisy data stream from a single LIGO detector. Since this is a time-dependent process, we need to ensure that SMTDAE  can recover a signal given noisy signal input and return zeros given pure noise.
	
	Denoising GWs is similar to removing noise in automatic speech recognition (ASR) through RNN, as illustrated in FIG~\ref{MTDAE}. The state-of-the-art tool in ASR is the Multiple Timestep Denoising Autoencoder (MTDAE), introduced in~\cite{DBLP:conf}. The idea of this model is to take multiple time steps within a neighborhood to predict the value of a specific point. Compared to conventional RNNs, which takes only one time step input to predict the value of that corresponding output, MTDAE takes one time step and its neighbors to predict one output. It is shown in~\cite{DBLP:conf} that this model returns better denoised outputs. 
	
	Realizing the striking similarities between ASR and denoising GWs, we have constructed a Staired Multiple Timestep Denoising Autoencoder (SMTDAE). As shown in FIG~\ref{SMTDAE}, our new model encodes the actual physics of the problem we want to address by including the following novel features:
	
	\begin{itemize}
	\item Since GW detection is a time-dependent analysis, our encoder and decoder have time-correlations, as shown in FIG~\ref{SMTDAE}. The final state that records information of the encoder will be passed to the first state of the decoder. We use a sequence-to-sequence model~\cite{sutskever2014sequence} with two layers for the encoder and decoder, where each layer uses a bidirectional LSTM cell~\cite{hochreiter1997long}. This type of structure is widely used in Natural Language Processing (NLP)~\footnote{A practical implementation of NLP for LIGO was recently described in~\cite{1710.05350}}.
	\item We have included another scalar variable which we call Signal Amplifier---indicated by a green circle in FIG~\ref{SMTDAE}. This is extremely helpful in denoising GW signals when the amplitude of the signal is lower than that of the background noise. Specifically, we use 9 time steps to denoise inputs for one time step. For each hidden layer in the encoder and decoder, we have 64 neurons. 
	\end{itemize}
	
	The key experiments which we conducted and the results of our analysis are presented in the following sections.

	\section{Experiments}
	\label{exp}
	
	For this analysis, we use simulated gravitational waveforms that describe binary black hole (BBH) mergers, generated with the waveform model introduced in~\cite{bohe:2017}, which is available in LIGO's Algorithm Library~\cite{LAL}. We consider BBH systems with mass-ratios \(q\leq10\) in steps of 0.1, and with total mass \(M\in[5\msun, 75\msun]\), in steps of \(1\msun\) for training. Intermediate values of total mass were used for testing. The waveforms are generated with a sampling rate of 8192 Hz, and whitened with the design sensitivity of LIGO~\cite{ZDHP:2010}. We consider the late inspiral, merger and ringdown evolution of BBHs, since it is representative of the BBH GW signals reported by ground-based GW detectors~\cite{DI:2016,secondBBH:2016,thirddetection,fourth:2017}. We normalize our inputs (signal+noise) by their standard deviation to ensure that the variance of the data is 1 and the mean is 0. In addition, we add random time shifts, between 0\% to 15\% of the total length, to the training data to make the model more resilient to variations in the location of the signal. Only simulated additive white Gaussian noise was added during the training process, while real non-Gaussian noise, 4096s taken from the \href{https://losc.ligo.org/about/}{LIGO Open Science Center} (LOSC) around the LVT151012 event, was whitened and added for testing.

	Decreasing SNR over the course of training can be seen as a continuous form of transfer learning~\cite{Weiss2016}, called Curriculum Learning (CL)~\cite{bengio2009curriculum}, which has been introduced in~\cite{2017arXiv170100008G} for dealing with highly noisy GW signals. Signals with high peak SNR > 1.00 (MF SNR > 13) can be easily denoised, as shown in FIG~\ref{more_imgs}. When the training directly starts with very low SNR from the beginning, it is difficult for a model to learn the original signal structure and remove the noise from raw data. To denoise signals with extremely low SNR, our training starts with a high peak SNR of 2.00 (MF SNR = 26) and then it gradually decreases every round during training until final peak SNR of 0.50 (MF SNR = 6.44). 
	
	\section{Results}
\begin{figure*}
		\begin{center}
			\subfigure[ SMTDAE]{\label{real13_noise:img4}\includegraphics[width=0.325 \textwidth]{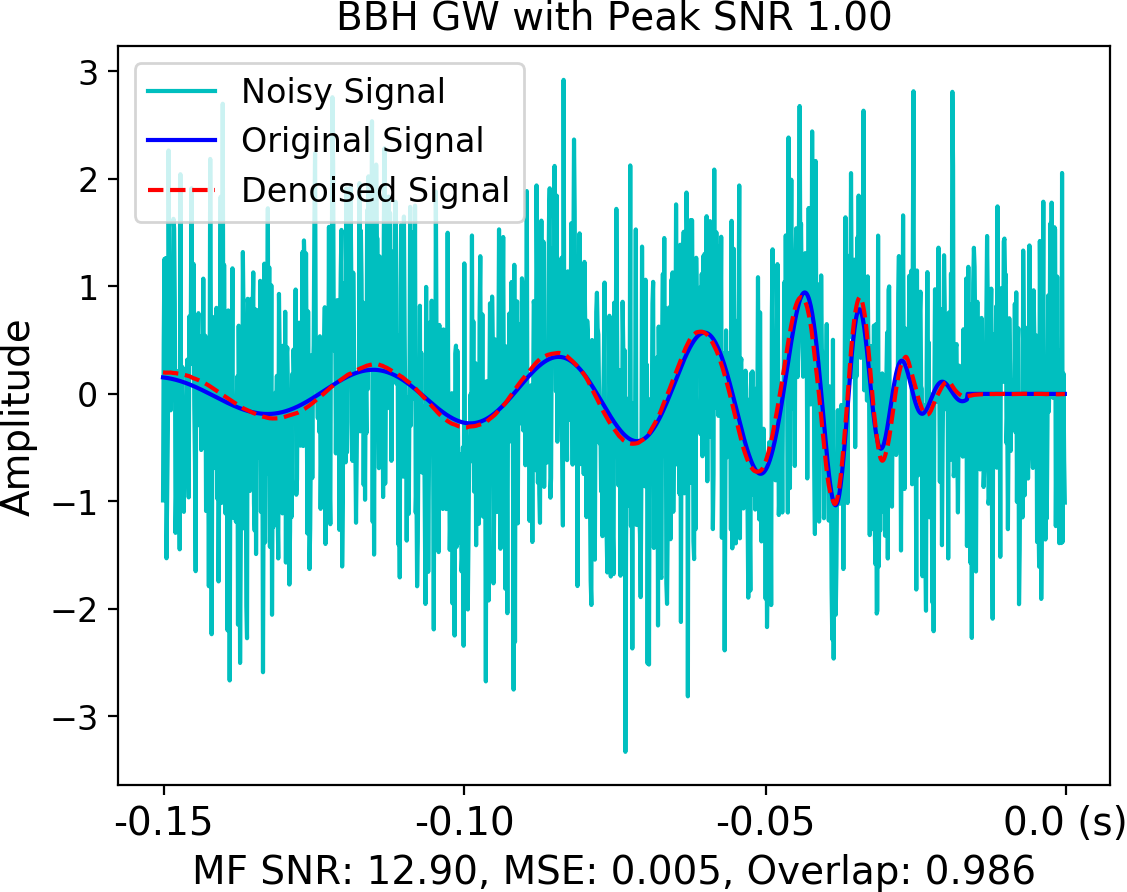}}
			\subfigure[ Dictionary Learning]{\label{real13_noise:img5}\includegraphics[width=0.325 \textwidth]{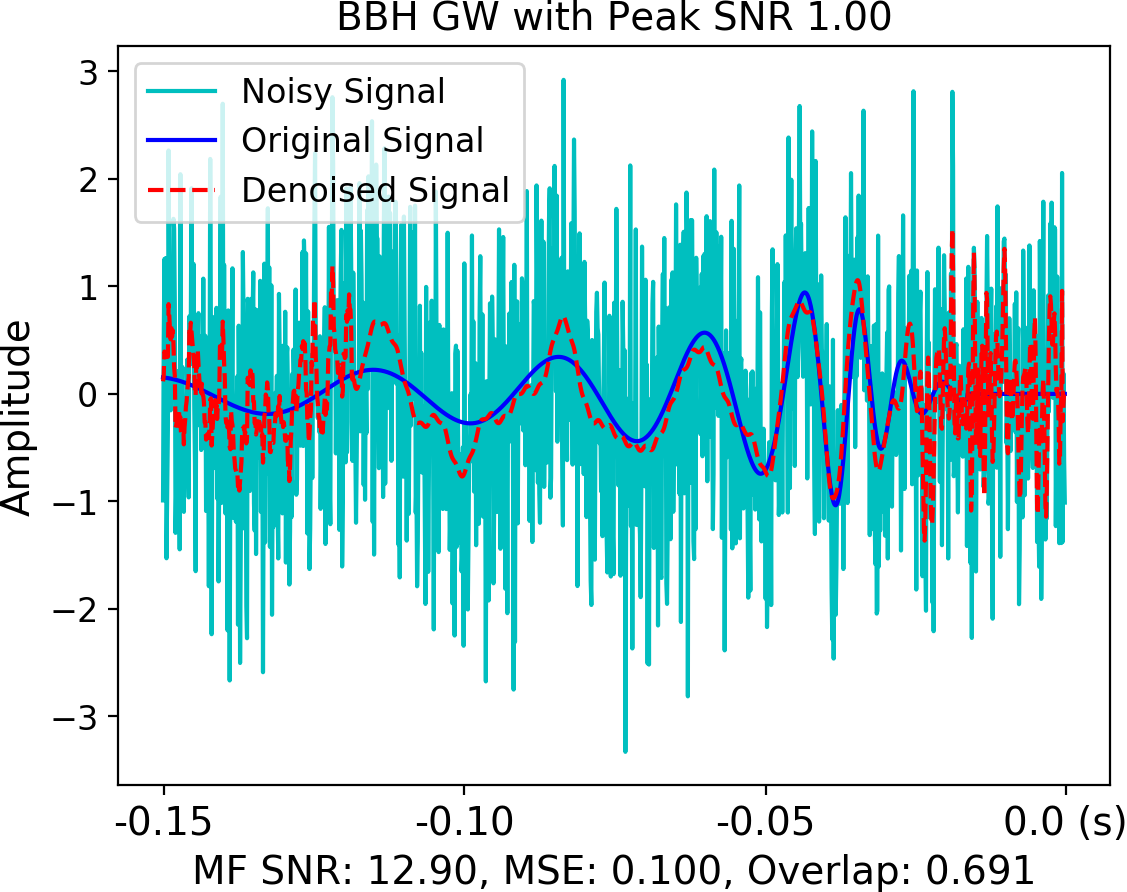}}
			\subfigure[ PCA]{\label{real13_noise:img6}\includegraphics[width=0.325 \textwidth]{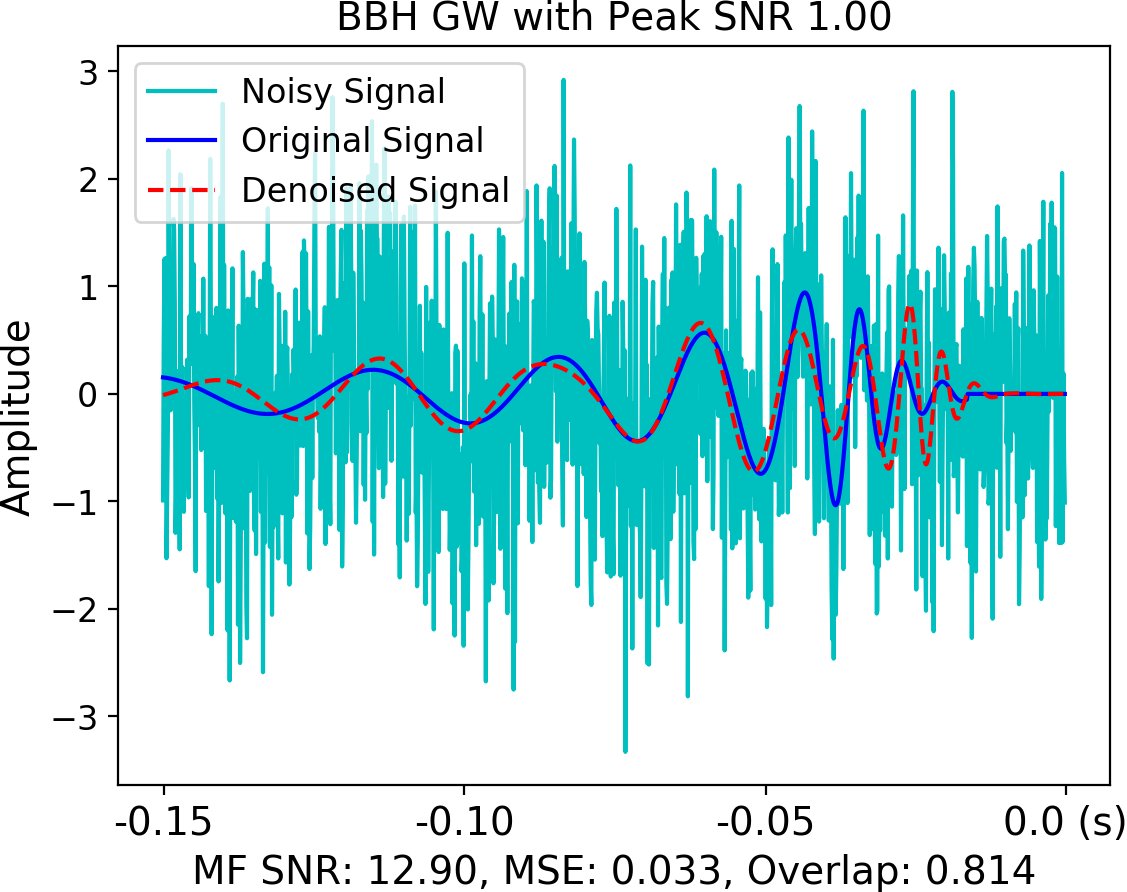}} \\
			\subfigure[ SMTDAE]{\label{real13_noise:img1}\includegraphics[width = 0.325 \textwidth]{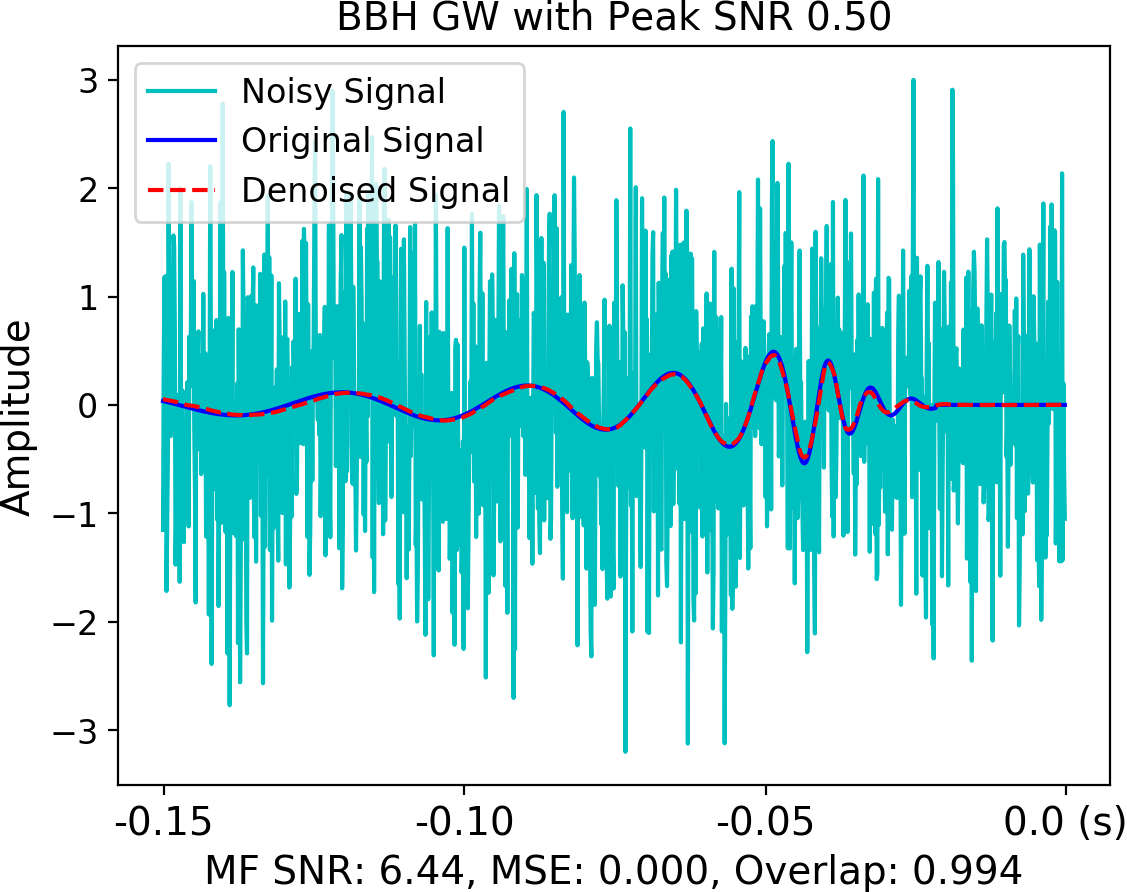}}
			\subfigure[ Dictionary Learning]{\label{real13_noise:img2}\includegraphics[width=0.325 \textwidth]{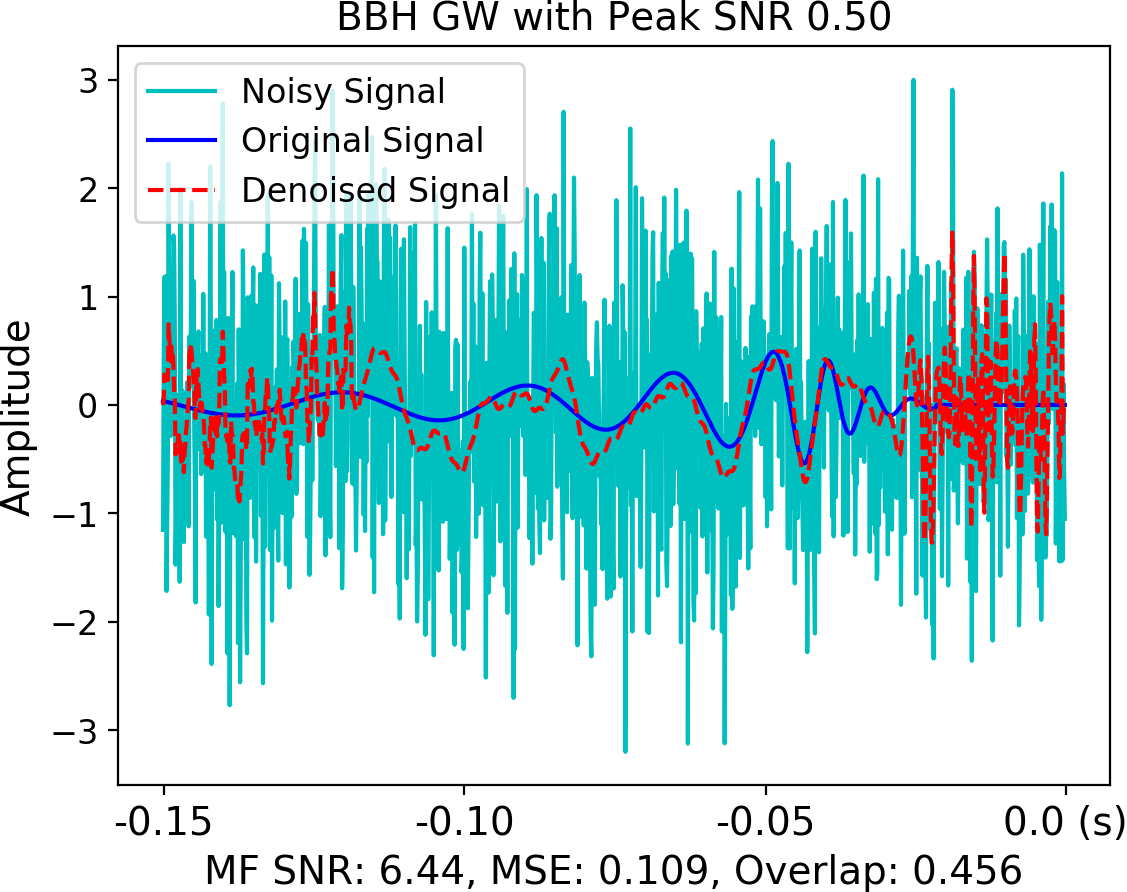}}
			\subfigure[ PCA]{\label{real13_noise:img3}\includegraphics[width=0.325 \textwidth]{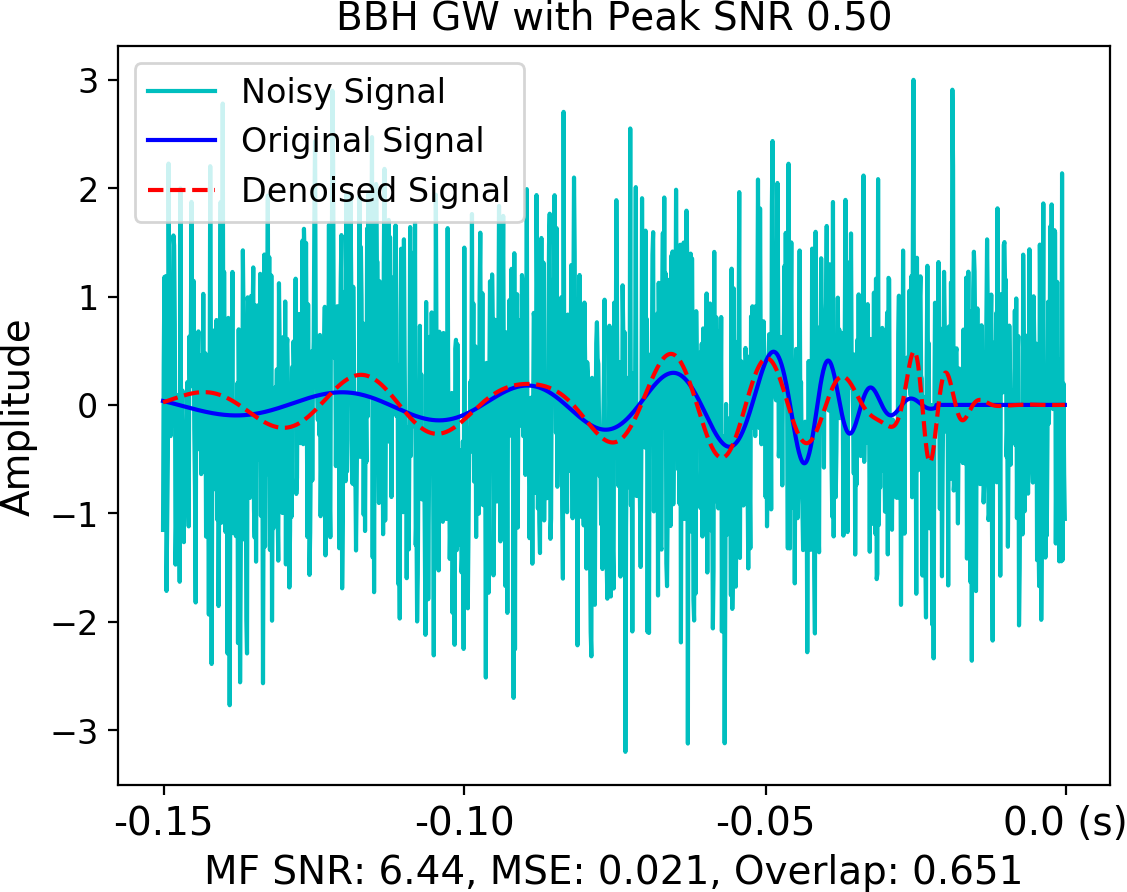}} \\
			\subfigure[ SMTDAE]{\label{real13_noise:img10}\includegraphics[width=0.325 \textwidth]{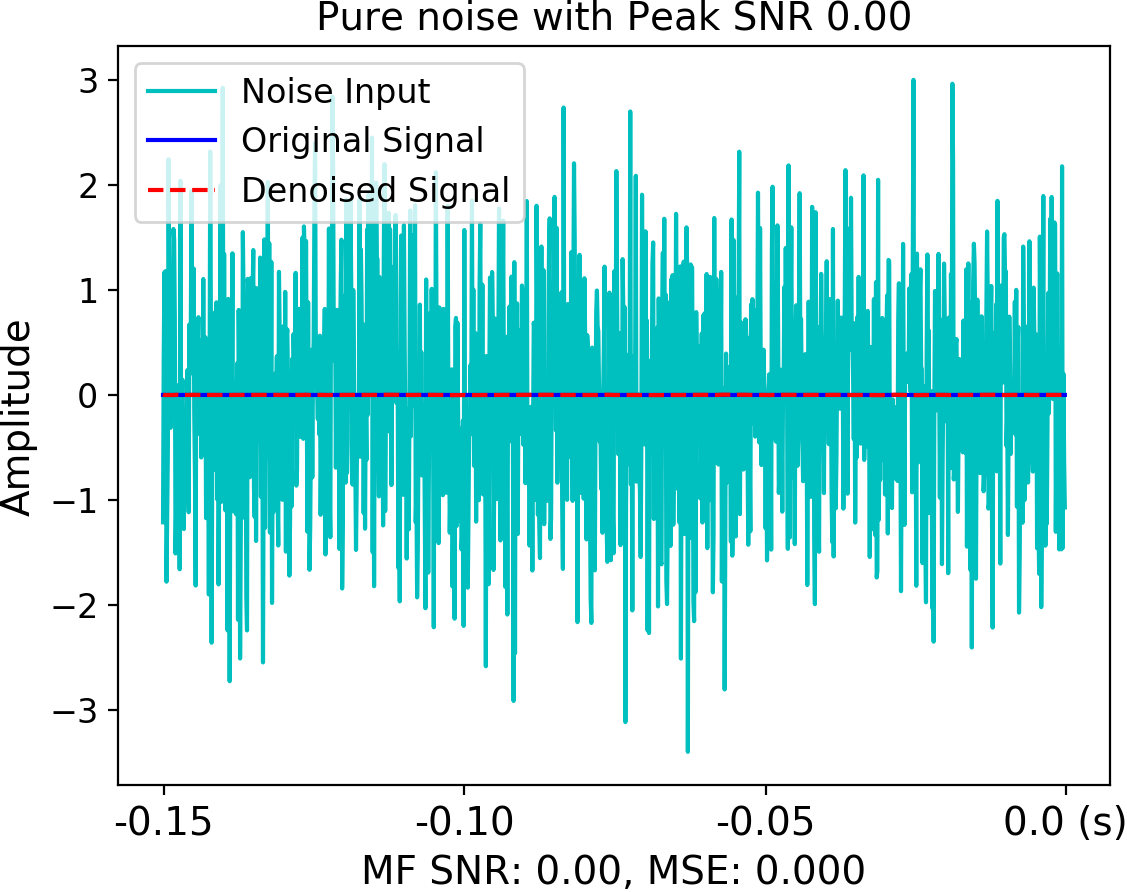}}
			\subfigure[ Dictionary Learning]{\label{real13_noise:img11}\includegraphics[width=0.32\textwidth]{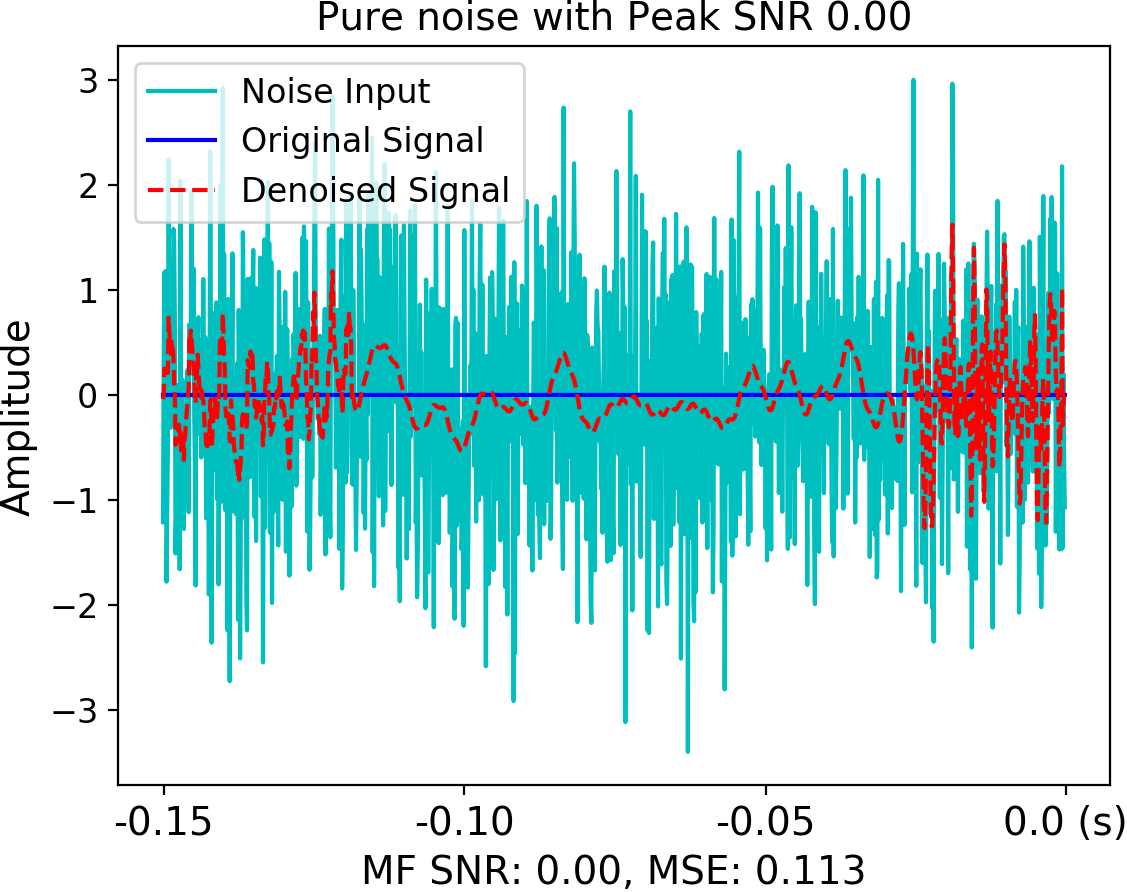}}
			\subfigure[ PCA]{\label{real13_noise:img12}\includegraphics[width = 0.325 \textwidth]{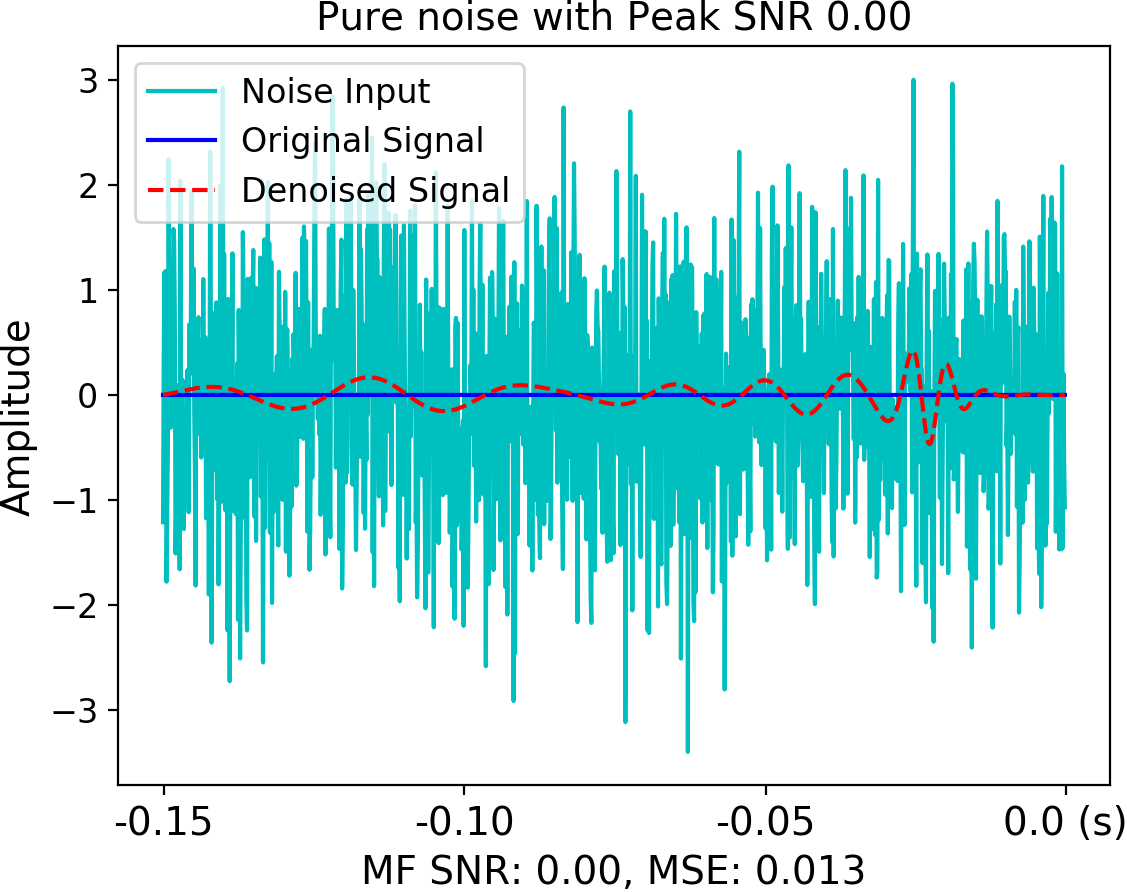}}
		\end{center}	
		\caption{Denoising results on test set signals injected into real non-Gaussian LIGO noise. \ref{real13_noise:img4}, \ref{real13_noise:img1} and \ref{real13_noise:img10} show results of SMTDAE trained only on simulated Gaussian noise on signals injected into real LIGO noise with peak SNR 0.50 and 1.00---equivalent to MF SNR of 6.44 and 12.90, respectively---and on pure LIGO noise (SNR 0.00).  \ref{real13_noise:img5}, \ref{real13_noise:img2} and \ref{real13_noise:img11} show corresponding results for dictionary learning model described in~\cite{PhysRevD.94.124040}. \ref{real13_noise:img6}, \ref{real13_noise:img3} and \ref{real13_noise:img12} show results for PCA model with 10 principal components. The length of each principal component is same as the length of a signal. Peak SNR, optimal matched-filtering SNR (MF SNR), mean square error (MSE) and Overlap are indicated in each panel.}
		\label{more_imgs}
		\vspace{-5mm}
	\end{figure*}

	All our training session were performed on NVIDIA Tesla P100 GPUs using TensorFlow~\cite{1603.04467}. We show the results of denoising with our model using signals from the test set injected into \textit{real} LIGO noise in FIG~\ref{more_imgs}, and compare them with PCA and dictionary learning methods (using the code based on~\cite{DictionaryLearningCode}). MSE and Overlap are reported with each figure. MSE is a measure of $L_2$ distance in vector space of GWs, whereas Overlap indicates the level of agreement between the phase of the two signals. Since both MSE and Overlap provide complementary information about the denoised waveforms, we include both measurements in our analysis. 
	
	In FIG~\ref{more_imgs}, we show results with PCA, dictionary learning, and SMTDAE, on the test set signals embedded in real LIGO noise. Note that our model was only trained with white Gaussian noise. We show that after training at different SNRs, our model outperforms PCA and dictionary learning in terms of the MSE and Overlap in the presence of real LIGO noise. In addition, our model is able to return a flat output of zeros when the inputs are either pure Gaussian noise or non-Gaussian, non stationary LIGO noise. In terms of computational performance, PCA takes on average two minutes to denoise 1s of input data. In stark contrast, applying our SMTDAE model with a GPU, takes on average less than 100 milliseconds to process 1s of input data.
	
	\section{Conclusion}
	\label{con}
	
	We have introduced SMTDAE, a new non-linear algorithm to denoise GW signals which combines a DAE with an RNN architecture using unsupervised learning. When the input data is pure noise, the output of the SMTDAE is close to zero. We have shown that the new approach is more accurate than PCA and dictionary learning methods at recovering GW signals in real LIGO noise, especially at low SNR, and is significantly more computationally efficient than the latter. More importantly, although our model was trained only with additive white Gaussian noise, SMTDAE achieves excellent performance even when the input signals are embedded in real LIGO noise, which is non-Gaussian and non-stationary. This indicates SMTDAE will be able to automatically deal with changes in noise distributions, without retraining, which will occur in the future as the GW detectors undergo modifications to attain design sensitivity.
	
	We have also applied SMTDAE to denoise new classes of GW signals from eccentric binary black hole mergers, simulated with the Einstein Toolkit~\cite{ETL:2012CQGra}, injected into real LIGO noise, and found that we could recover them well even though we only used non-spinning, quasi-circular BBH waveforms for training. This indicates that our denoising method can generalize to new types of signals beyond the training data. We will provide detailed results on denoising different classes of eccentric and spin-precessing binaries as well as supernovae in a subsequent extended article. The encoder in SMTDAE may be used as a feature extractor for unsupervised clustering algorithms~\cite{DeepTransferNIPS}. Coherent GW searches may be carried out by comparing the output of SMTDAE across multiple detectors or by providing multi-detector inputs to the model. Denoising may also be combined with the Deep Filtering technique~\cite{2017arXiv170100008G,DNN2NIPS} for improving the performance of signal detection and parameter estimation of GW signals at low SNR, in the future. We will explore the application of this algorithm to help detect GW signals in real discovery campaigns with the ground-based detectors such as LIGO and Virgo.

	\bibliography{references}

\end{document}